\documentclass[10   xpt,conference]{IEEEtran}
\IEEEoverridecommandlockouts
\usepackage{cite}
\usepackage{amsmath,amssymb,amsfonts}
\usepackage{algorithmic}
\usepackage{graphicx}
\usepackage{comment}
\usepackage{textcomp}
\usepackage{xcolor}
\usepackage{url}
\usepackage[T1]{fontenc}
\def\BibTeX{{\rm B\kern-.05em{\sc i\kern-.025em b}\kern-.08em
    T\kern-.1667em\lower.7ex\hbox{E}\kern-.125emX}}
    
\begin{document}

\title{Towards Greener Data Centers via Programmable Data Plane}



\author{
 Garegin Grigoryan\\ grigoryan@alfred.edu  \\Alfred University 
 \and
 Minseok Kwon \\ mxkvcs@rit.edu \\Rochester Institute of Technology 
}

\maketitle
\begin{abstract}
The energy demands of data centers are increasing and are expected to grow exponentially. Reducing the energy consumption of data centers decreases operational expenses, as well as their carbon footprint.
We design techniques to reduce data center power consumption by leveraging Software-Defined Networking (SDN) and programmable data plane concepts. Relying solely on in-data plane registers, our proposed system P4Green consolidates traffic in the least number of network switches and shifts workloads to the servers with the available renewable energy. Unlike existing SDN-based solutions, P4Green's operation does not depend on a centralized controller, making the system scalable and failure-resistant. Our proof-of-concept simulations show that traffic consolidation can reduce data centers' aggregation switch usage by 36\% compared to standard data center load balancing techniques, while workload control can boost renewable energy consumption for 46\% of the daily traffic.
\end{abstract}

\section{Introduction}
\label{sec:intro}
With the rapid growth of cloud-based data aggregation and computation, the data centers hosting clouds became the major consumers of electrical power and produce a significant carbon footprint~\cite{siddik2021environmental, engie}. A Data Center Network (DCN) that includes servers and network infrastructure consumes about 40-50\% of the total energy consumed by a data center overall, and that share is likely to grow~\cite{rong2016optimizing, bilal2014taxonomy}. As the efficiency of data centers' non-IT components improves~\cite{googlecooling, masanet2020recalibrating}, optimizing the energy consumption of computing and networking components becomes critical. Only by 2030, data center user behavior is expected to increase the energy consumption of data centers by 20\% in case the current technological development trends do not change. However, the likely end of Moore's law may lead to data centers' energy consumption surge by up to 134\%~\cite{koot2021usage}.

Energy consumption of network components is not proportional to traffic~\cite{frohlich2021smart}, hence even switches that transmit little traffic consume considerable energy resources. To reduce the energy consumption of network infrastructure, the state-of-the-art approaches propose carefully choosing the architecture for a data center (e.g., three-tier vs fat-tree data center network) to maximize the utilization of the available resources in a data center. To reduce the energy consumption of servers, workload scheduling systems~\cite{grange2018green}, virtualization techniques~\cite{helali2021survey}, and server hardware optimizations~\cite{katal2022energy} were proposed. Using renewable energy can reduce the carbon footprint of data center servers. One challenge is the volatility of renewable energy and the high costs of its storage~\cite{pei2020minimal}. 
To overcome this, workload migration techniques were proposed~\cite{siddik2021environmental, zheng2020mitigating}. Often these approaches require expensive hardware modifications; in the meantime, re-organizing the data center and consolidation of resources may lead to a lack of redundancy and therefore decreased quality of service during peak hours of a data center operation. 

In this work, we present \textit{P4Green}, a dynamical adaptive system aimed at reducing the energy consumption of server and network components at a data center. Instead of relying on dedicated hardware, we leverage the technologies of switch programmability. Offloading green traffic engineering to programmable switches has several benefits. First, such switches can be programmed to arbitrarily process each data packet at the line rate; second, they can collect and store information reflecting the state of the data center network and its servers, including their renewable energy capacities. In addition, programmable switches are faster and more power-efficient than switches with fixed-function ASIC~\cite{agrawal2020intel, arista}. 

Specifically, we leverage the Software-Defined Networking (SDN) via P4 programming language for the data plane and P4Runtime API for the control plane~\cite{bosshart2014p4, p4runtime}. The goal of P4Green is to consolidate traffic in the least number of switches based on the traffic volume and to schedule workloads on servers with the available renewable energy. P4Green switches evaluate the data center traffic per pre-defined time epochs and based on that adjust the packet forwarding algorithms to include or exclude additional network switches. In addition, they collect information about the availability of renewable energy at the data center servers, make workload allocation decisions, and forward the traffic to the most appropriate server. Unlike existing SDN-based traffic engineering solutions, P4Green does not rely on a controller and uses only its in-data plane registers for making forwarding decisions. P4Green operates in the data plane at the line rate, improving the scalability of the system and making it robust to control plane failures. 




To summarize our contributions, we:

\begin{itemize}
    \item Designed P4Green, a system for reducing power consumption of network devices in a data center and load balancing workloads towards servers with available green resources;
    \item P4Green traffic engineering and forwarding operates at the line rate, fully in the data plane; the control plane is used only at the initialization stage;
    \item We implemented and tested the prototype of P4Green using Mininet and bmv2 switch emulator;
    \item Our results show a 36\% reduction of data centers' aggregation switch usage compared to a traditional ECMP load balancing; as well as a boost of energy consumption for 46\% of the daily traffic in a distributed data center scenario.
\end{itemize}

\section{Design}
\begin{figure*}
	\begin{center}
		\includegraphics[width=1.7\columnwidth]{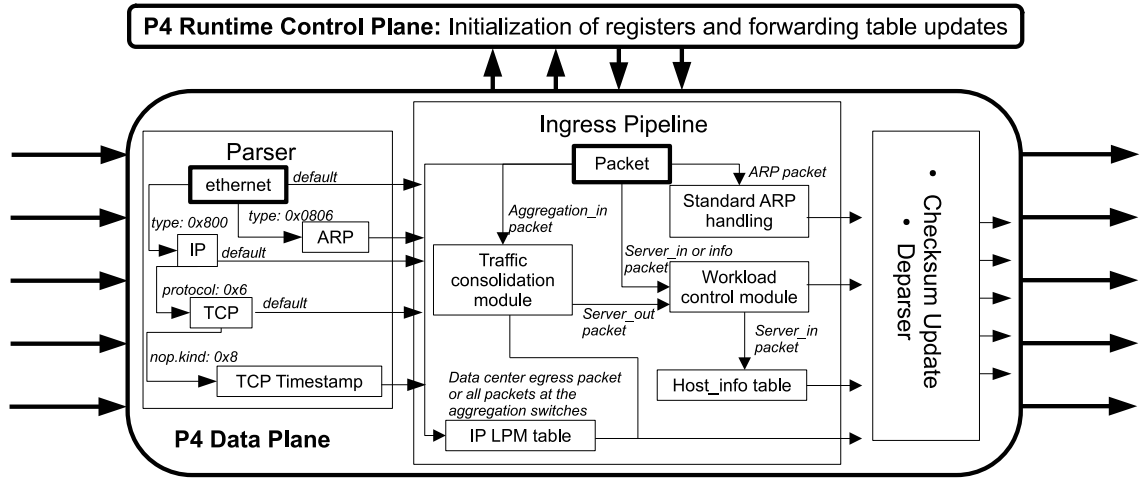}
		\caption{\label{fig:architecture}P4Green architecture and workflow}

	\end{center}
\end{figure*} 
\label{sec:design}
P4Green leverages the properties of the P4 switch architecture that allow limited programmability in the forwarding engine and full programmability at the control plane. The goals of P4Green are:
\begin{itemize}
    \item \textbf{Traffic consolidation during light load.}
    P4Green consolidates traffic by dynamically changing the number of active aggregation switches based on traffic volume, with no control plane involvement.
    \item \textbf{Green energy-driven workload control}. Server workload is determined by the resources available at the server such as renewable energy. Switches obtain resource information from each server, choose appropriate servers for target workloads, and forward workload traffic to the chosen servers.
\end{itemize}

Overall, P4Green helps reduce the energy consumption of a data center network by minimizing the number of active switches; in addition, P4Green boosts renewable energy usage by sending workloads toward servers that report the availability of such energy. In this section, we describe how we implement the functionality of P4Green while overcoming challenges, such as the data plane computational limitations, the dependency on the control plane, and the TCP session affinity requirements.

\subsection{Overview} The architecture of P4Green is shown in Figure~\ref{fig:architecture} with its workflow. P4Green is implemented with a single P4 program for the data plane and a Python program for the SDN controller that uses P4Runtime API to set up the switches. 
When a packet arrives at the switch, it first is parsed to extract information needed for packet forwarding and future analysis. In addition to the standard fields such as Ethernet and IP headers, the parser extracts the TCP header as well as the TCP timestamp option. The TCP header is for load balancing across different aggregation switches via 5-tuple hashing. The TCP timestamp option enables TCP session affinity when distributing traffic to different servers. After parsing, the packet is passed to the ingress pipeline in which an output port is assigned based on the following factors:
\begin{itemize}
    \item Switch type: \textit{core}, \textit{aggregation}, or \textit{access}; 
    \item Packet type: (a) \textit{Aggregation\_in}, packets that move from non-aggregation switches towards aggregation switches; (b) \textit{Server\_in}, packets that move in to servers; (c) \textit{Server\_out}, packets that move out from servers.
\end{itemize}
As shown in Figure~\ref{fig:architecture}, the ingress pipeline consists of the \textit{Traffic Consolidation} and \textit{Workload Control} modules together with \textit{Longest Prefix Matching} (LPM) and the \textit{Host\_info} match-action tables. Next, we describe these components in detail.
\begin{figure}
	\begin{center}
		\includegraphics[width=0.9\columnwidth]{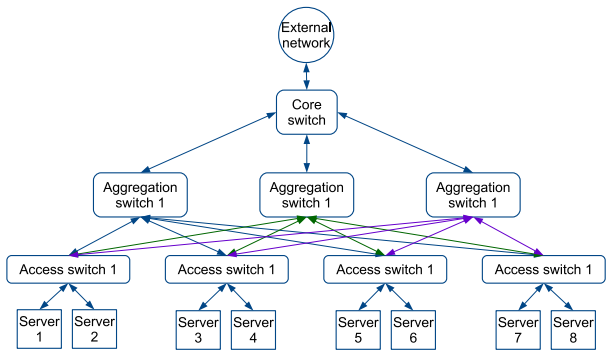}
		\caption{\label{fig:mininet}A simplified data center network}
	\end{center}
 \vspace{-5mm}
\end{figure}

\begin{figure}
	\begin{center}
		\includegraphics[width=1\columnwidth]{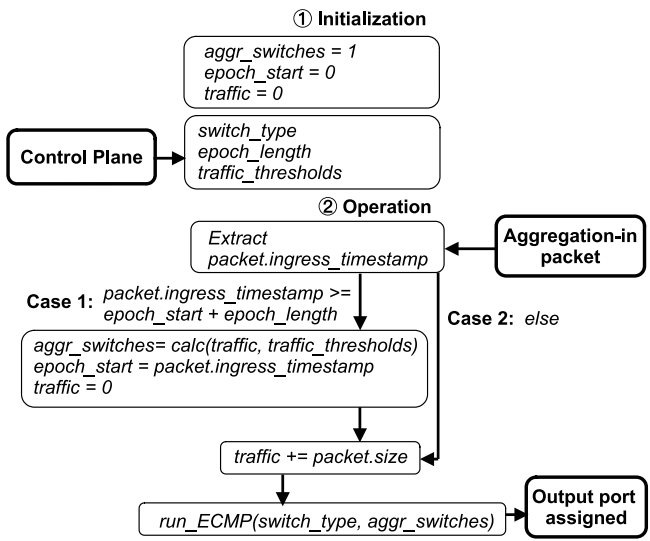}
		\caption{\label{fig:trafficconsol}Traffic Consolidation workflow}
  \vspace{-4mm}
	\end{center}
\end{figure}

\subsection{Traffic Consolidation}
A typical data center network has three layers of switches: \textit{access switches} that connect servers to the network, \textit{core switches} that forward heavy inter-traffic, and \textit{aggregation switches} that forward inter-traffic between access and core switches as well as intra-traffic between access switches (see Figure~\ref{fig:mininet}). The connectivity between the access and core layers exists as long as there is at least one active aggregation switch. The \textit{Traffic Consolidation} module helps determine the number of aggregation switches to be activated dynamically depending on traffic volume. Specifically, it estimates traffic arriving at core and access switches per epoch time, enables (or disables) more aggregation switches to forward traffic if traffic volume grows (or drops) significantly. This module runs only for ingress packets at core switches (\textit{Aggregation\_in} packets), and for both ingress and server packets at access switches (\textit{Aggregation\_in} and \textit{Server\_out packets}). 
The workflow of the module is defined in more detail in  Figure~\ref{fig:trafficconsol} and in the following subsections.
\subsubsection{Initialization} First, three registers are initialized:
\begin{itemize}
    \item \textit{aggr\_switches}: number of enabled switches (initially set to 1).
    \item \textit{epoch\_start}: starting timestamp for measuring traffic within a time interval.
    \item \textit{traffic}: counter to estimate traffic volume.
\end{itemize}
These registers are set to the same values for all switches. This can be done solely at the switches themselves with no controller involved. There are also other registers possibly having different values based on the network topology and the preferences of network operators. They are indeed to be initialized by the controller:
\begin{itemize}
    \item \textit{switch\_type}: Core or access switches execute switch-type-specific commands while aggregation switches simply forward packets to the destination IP address (see Figure~\ref{fig:architecture}).
    \item \textit{epoch\_length}: The length of an epoch is used for estimating traffic volume.
    \item \textit{traffic\_threshold(s)}: This threshold(s) is used to enable or disable aggregation switches by recalculating the \textit{aggr\_switches} register.
\end{itemize}
Note that these registers can be modified during the operation of the switch using either the control or the data plane. As we show further in this section, some of these registers are modified automatically by the data plane of the switch.

\subsubsection{Operation} Core and access switches use ECMP for  forwarding packets to aggregation switches based on 5-tuple hashing. These switches use \textit{aggr\_switches} as the width of the ECMP hashing, i.e., the number of possible distinct outputs, which controls the number of aggregation switches that forward traffic. The value of \textit{aggr\_switches} changes based on the traffic volume that core and access switches process. The counter \textit{traffic} is updated for each packet arrival and reset to zero at the end of each epoch, defined by  \textit{epoch\_length}. All updates to \textit{traffic} are done directly in the switch.


Specifically, for each incoming packet, the P4Green program reads the ingress timestamp in the metadata. If the timestamp exceeds \textit{epoch\_start} by \textit{epoch\_length}, then   \textit{traffic} contains the traffic volume the switch has processed during \textit{epoch\_length}. The traffic volume is then used to recalculate \textit{aggr\_switches} by comparing \textit{traffic} to \textit{traffic\_thresholds} provided by the control plane. Depending on the measured traffic volume, \textit{aggr\_switches} is either increased or decreased, and then used as the width of the ECMP hashing for assigning output ports. Finally, \textit{traffic} is reset to zero and \textit{epoch\_start} is reset to the current timestamp, and traffic load evaluation is restarted for a new epoch.

In P4Green, the initialization is the only step that needs input from the control plane. The control plane program uses P4Runtime API to initialize the switch state variables. The traffic load evaluation and ECMP modification are performed solely in the data plane. In the meantime, P4Green requires a power control module to power off switches that do not forward any traffic. Such a module can control the switches' state by checking the value in \textit{aggr\_switches} register of the switch. We leave the design of the power control module out of the scope of this paper.


\subsection{Workload Control}
\begin{figure}
	\begin{center}
		\includegraphics[width=1\columnwidth]{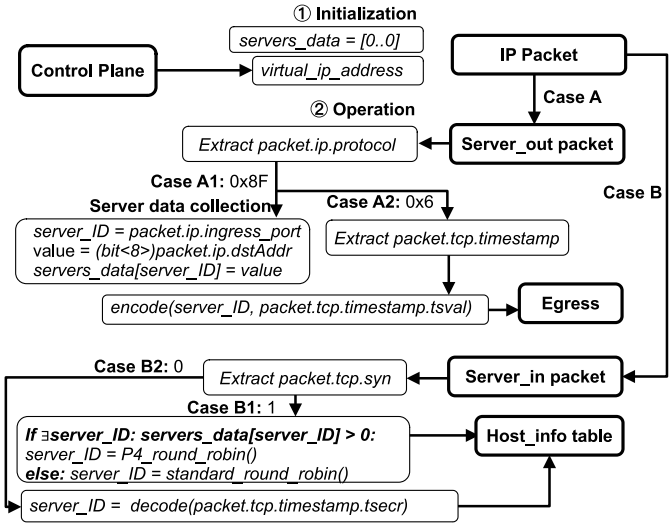}
		\caption{\label{fig:workloadctrl}Workload Control workflow}
  \vspace{-6mm}
	\end{center}
\end{figure}

Distributed data centers naturally have heterogeneous servers in terms of geographical locations, renewable energy and volatile resources. In iLoad~\cite{grigoryan2019iload}, servers determine their workload, e.g., machine learning tasks or web traffic, based on locations and resources. A programmable switch receives those workload requests directly from servers, and makes adjustments to packet forwarding but with little input from the controller. The controller in iLoad regularly updates the forwarding table at the switch, using the servers' workload requests information stored in the data plane.
P4Green adopts a similar approach, but enhances it by completely eliminating the control plane from switch operation. We also use the TCP timestamp to provide TCP session affinity, leveraging the approach presented in~\cite{barbette2020high}. The algorithm is discussed below with its workflow depicted in Figure~\ref{fig:workloadctrl}.

\subsubsection{Initialization} 
\label{workloadctrl}

At the startup, registers and tables are initialized:
\begin{itemize}
    \item The \textit{servers\_data} register block for storing per-server resource availability index is initially set to 0 for every server. \textit{servers\_data} is used to select the most pertinent server ID for the target workload.
    \item The \textit{virtual\_ip} address, set by the control plane for processing client packets.
\end{itemize}
In addition, the switch has the \textit{Host\_info} table with actions to be performed when forwarding packets based on the server's ID (i.e., rewriting the MAC headers and the destination IP address).

\subsubsection{Operation} First, P4Green classifies the incoming packets into \textit{Server\_out} (Case A) and \textit{Server\_in} (Case B) categories. For each category, further classification is performed by P4Green.

\textit{a) Info-packets:} A server sends the switch an info-packet that contains a resource availability index (e.g., renewable energy, CPU, or memory availability). 
An info-packet is an IP packet with the first three octets of the destination IP address taken from the subnet to which the switch belongs and the last octet used to encode the payload.  The protocol field has the specific value (0x8F in our example) for the switch to identify info-packets. 

Handling info-packets is illustrated on Figure~\ref{fig:workloadctrl}, Case A1. The P4 program parses info-packets to obtain the sender ID (ingress port plus source IP address) and the resource availability index. The index is stored in \textit{server\_data} and used to choose the most relevant server for client workload requests.

\textit{b) Client requests:} Each access switch is assigned with a Virtual IP address (VIP) at the initialization time, and clients send \textit{Server\_in} packets using VIPs. Assuming the data center uses the TCP protocol, upon receiving a \textit{Server\_in} packet, the switch extracts the TCP SYN flag  to identify if it is a new request (\textit{Case B1} in Figure~\ref{fig:workloadctrl} when SYN=1). The switch then selects the most relevant server to handle this client request. 
If no server has reported the resource availability index greater than zero, the switch simply uses round-robin to select the server. If there are such servers, the switch selects the next server with available resources. The packet is then augmented with the server ID and passed to the \textit{Host\_info} match-action table that in turn replaces the VIP with the destination IP address of the selected server.

If SYN=0 (\textit{Case B2} in Figure~\ref{fig:workloadctrl}), the server that previously served the corresponding TCP session needs to be selected for the TCP session affinity requirement. Finding such a server is, however,  challenging since the original destination IP address is the VIP, not the server IP address. In the meantime, storing a per-session table of connections in the switches is not scalable. To get around this problem, we leverage TCP option fields to encode the server ID for each flow as proposed in~\cite{barbette2020high}. It uses the TCP property that hosts echo the TCP timestamp (in \textit{ms}) to each other while in session. A TCP \textit{Server\_out} packet encodes the server ID into the last three bits of timestamp (see Case A2 for \textit{Server\_out} packets in Figure~\ref{fig:workloadctrl}). Now, the server ID can be obtained from the last three bits of the TCP timestamp echo field for \textit{Seriver\_in} packets of existing client requests (when SYN$=$0). The packet is then matched to the \textit{Host\_info} table to rewrite its headers. Note that we use only the three least significant bits of the TCP timestamp here, while the rest is not affected. Our solution does not require modifications to the server or client TCP/IP stack, except enabled TCP timestamp support. An alternative to leveraging TCP timestamp is using other reliable protocols such as QUIC~\cite{langley2017quic} with the \textit{Connection ID} field in its header.

\section{Evaluation}
\label{sec:eval}
\begin{figure}
	\begin{center}
		\includegraphics[width=0.9\columnwidth]{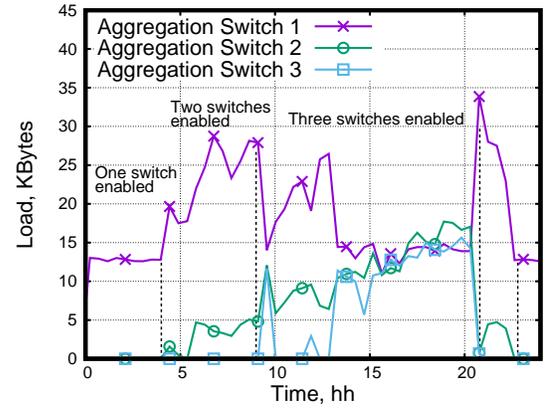}
  \vspace{-5mm}
		\caption{\label{fig:consolidation}Traffic consolidation in P4Green}
 \vspace{-3mm}
	\end{center}
\end{figure} 

We implement P4Green as a Mininet emulator on the bmv2~\cite{bmv2} switch, and test its traffic consolidation and workload control aspects as a proof-of-concept. For testing, we use the network topology illustrated in Figure~\ref{fig:mininet} that consists of eight servers with unique IP addresses, four access switches with virtual IP addresses, three aggregation switches, and one core switch connected to the external network. First, traffic is sent from outside to the access switches, and we measure how traffic volume affects packet forwarding. Moreover, we test how traffic consolidation helps minimize network resources usage. We use iperf3~\cite{iperf} for traffic generation. After that, Hosts 1 and 2 send info-packets to access switch 1 with their renewable energy availability index. An external host then sends workload requests to VIP 1, and we observe which servers respond to those requests.

Note that the Mininet emulator used in our experiments is not designed for processing large volumes of traffic. Hence, the traffic rates in our evaluation emulate the patterns of the daily traffic and do not represent the realistic data center traffic volumes.


\subsection{Traffic Consolidation}

At access and core switches, we set the threshold for enabling two and three aggregation switches to 10KB and 20KB, respectively.
Traffic is measured every \textit{epoch\_length} which is set to 1 second. 
Traffic is generated with iperf3 to simulate a typical daily load with the peak in the afternoon. 
The results are displayed in Figure~\ref{fig:consolidation}. With the low traffic (approximately less than 20KB on Aggregation Switch 1 or $\approx$37\% of the peak load), only one aggregation switch receives traffic. As traffic grows, the second aggregation switch is activated in ECMP and helps decrease the load on the first switch. Once the traffic  exceeds a threshold of 32KB ($\approx$ 60\% of the peak load), the traffic flows through all aggregation switches. Overall, the operation hours of aggregation switches are reduced by more than 36\%. In a data center environment with up to 32 aggregation switches, each of them consuming about 400W*h (following the estimations made in~\cite{popoola2018energy}), that can save up to 4.6kW*h, not considering the reductions in the energy needs of cooling facilities due to reduced heat dissipation. The exact amounts of energy savings depend on the topology of the data center network and and its traffic behaviour.

\subsection{Workload Control}
\begin{figure}
	\begin{center}
		\includegraphics[width=0.9\columnwidth]{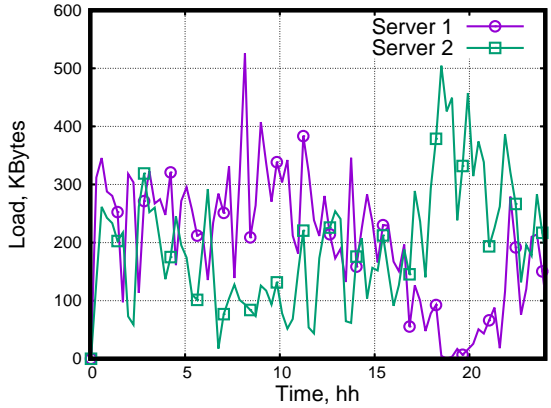}
		\caption{\label{fig:lb_nocontroller}Green load balancing in P4Green  (in Server 1's time zone)}
	\end{center}
\end{figure} 
\begin{figure}
	\begin{center}
		\includegraphics[width=0.9\columnwidth]{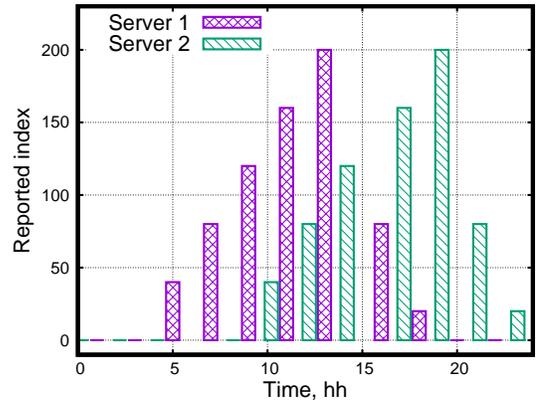}
		\caption{\label{fig:infodata}Reported renewable energy availability index by Server 1 and Server 2 (in Server 1's time zone)}
	\end{center}
\end{figure}
To test the workload control capabilities, we measure how traffic that servers receive changes as each server informs the closest access switch of available renewable energy. In the experiment, we simulate the case where Server 1 has more renewable energy in the first half of the day, while Server 2 has more in the second half. We assume that Server 1 is located in a time zone six hours ahead of the time zone of Server 2.

For the results, Figure~\ref{fig:lb_nocontroller} shows the traffic processed on Servers 1 and 2 during the day in the timezone of Server 1. The reported energy values follow the regular solar energy pattern on a sunny day (see Figure~\ref{fig:infodata}). As we simulate a distributed data center in different time zones, we assume the same overall traffic load throughout the day with all the hours presented in Server 1's time zone. Figure~\ref{fig:lb_nocontroller} shows that when both servers report zero renewable energy between 00h:00m and 05h:00m, the standard round-robin distributes the client workload evenly between the two servers. When Server 1 reports energy and Server 2 does not, the modified round robin is used to send more workload to Server 1. The figure shows that Server 1 processes a majority of the workload between 05h:00m and 13h:00m ($\approx$ 69\% of the total load). After the energy peak hours for Server 1 before 22h:00m, $\approx$ 68\% of the load is forwarded to Server 2 with more renewable energy during that time interval. Finally, at the end of the day (after 22h:00m), the load is distributed evenly, since both Server 1 and Server 2 are past their peak energy hours. In summary, P4Green distributes 46\% of the total traffic towards the servers that report more renewable energy. These results demonstrate that P4Green is successful in load balancing and traffic engineering with no help from control plane during the operational stage. 
Our implementation is publicly available\footnote{Link to the repository: \url{https://github.com/gareging/p4green}}.




\section{Related work}
\label{sec:related}
ElasticTree~\cite{heller2010elastictree} aggregates flows in the least number of networking devices thus reducing energy consumption. The ElasticTree controller periodically polls information from the data plane and updates the forwarding tables of the data plane. Li et al. in~\cite{li2014software} propose a different approach, allocating networking resources to flows exclusively, rather than letting them share the same links and switches. Such a design requires the control plane to analyze the packets of unseen flows which creates a risk of a control plane bottleneck. GRASP~\cite{grigoryan2018grasp} is designed with an SDN controller that subscribes to energy reports from the distributed data center servers and then re-actively installs forwarding rules for new flows into the data plane. As with~\cite{li2014software}, such an approach may not be scalable for a data center with large volumes of traffic. Chuang et al. in~\cite{chuang2021bandwidth} take into consideration a task execution length and use an OpenFlow-based controller to monitor and reschedule the data center jobs.  iLoad~\cite{grigoryan2019iload} is an in-network green load balancing solution; however, it still requires the control plane to analyze and update the data plane registers during switch operations. Xu et al. in~\cite{xu2017bandwidth} design a green scheduler for data center flows with deadlines. Work such as~\cite{cioara2015data, gao2020smartly} aimed at predicting green energy generation at data center servers and proactively scheduling workloads to the most active data centers.  The authors in~\cite{grange2018green} propose a green energy-aware algorithm for scheduling tasks in a data center.

\section{Conclusion}
We leverage the emerging technologies of programmable switches to design a data-plane system called P4Green that can allocate and de-allocate IT resources based on traffic load and renewable energy available at servers. P4Green eliminates the control plane from its operation and is shielded from control plane delays and failures. Our proof-of-concept implementation shows that it significantly reduces network switch usage and effectively allocates workloads to servers that report the availability of volatile renewable energy resources.

\bibliographystyle{IEEEtranS}
\bibliography{sample-base}
\end{document}